\documentclass[%
reprint,
superscriptaddress,
amsmath,amssymb,
pra,
]{revtex4-1}

\usepackage{graphicx}
\usepackage{dcolumn}
\usepackage{bm}
\usepackage{multirow} 

\usepackage[separate-uncertainty = true]{siunitx}
\usepackage[Symbol]{upgreek} 

\newcommand{\RN}[1]{\uppercase\expandafter{\romannumeral#1}}

\begin{document}

\title{Towards practical mass spectrometry with nanomechanical pillar resonators by surface acoustic wave transduction}

\author{Hendrik K\"ahler}
\affiliation{Institute of Sensor and Actuator Systems, TU Wien, Gusshausstrasse 27--29, 1040 Vienna, Austria.}
\author{Robert Winkler}
\affiliation{Christian Doppler Laboratory for Direct-Write Fabrication of 3D Nanoprobes (DEFINE), Institute of Electron Microscopy and Nanoanalysis, Graz University of Technology, Steyrergasse 17, 8010 Graz, Austria.}
\author{Holger Arthaber}
\affiliation{Institute of Electrodynamics, Microwave and Circuit Engineering, TU Wien, Gusshausstrasse 25, 1040 Vienna, Austria.}
\author{Harald Plank}
\affiliation{Christian Doppler Laboratory for Direct-Write Fabrication of 3D Nanoprobes (DEFINE), Institute of Electron Microscopy and Nanoanalysis, Graz University of Technology, Steyrergasse 17, 8010 Graz, Austria.}
\affiliation{Institute of Electron Microscopy and Nanoanalysis, Graz University of Technology, Steyrergasse 17, 8010 Graz, Austria.}
\affiliation{Graz Centre for Electron Microscopy, Steyrergasse 17, 8010 Graz, Austria.}
\author{Silvan Schmid}
\affiliation{Institute of Sensor and Actuator Systems, TU Wien, Gusshausstrasse 27--29, 1040 Vienna, Austria.}
\email{silvan.schmid@tuwien.ac.at}

\date{\today}

\begin{abstract}
Nanoelectromechanical systems (NEMS) have proven outstanding performance in the detection of small masses down to single proton sensitivity. To obtain a high enough throughput for the application in practical mass spectrometry, NEMS resonators have to be arranged in two-dimensional (2D) arrays. However, all state-of-the-art electromechanical transduction methods rely on electrical lines placed close to the mechanical resonators, which drastically restricts the density of 2D resonator arrays. An exception is the transduction by surface acoustic waves (SAWs), which has so far only been shown for the transduction of single nanomechanical pillar resonators. Here, we demonstrate the transduction of pillar pairs by surface acoustic waves. The pillars have a diameter of \SI{700}{\nano\meter} and show a mass responsivity of \SI{-588\pm98}{\nano\gram^{-1}}. The distances between the pillar pairs are \SI{70}{\nano\meter} and \SI{14.3}{\micro\meter}. SAW transduction enabled us to measure both pillars of each pair with electrical lines no closer than \SI{300}{\micro\meter}, illustrating the potential of SAWs to transduce dense arrays of pillar resonators, a crucial step towards practical mass spectrometry with NEMS. 


\end{abstract}

\maketitle

\section{Introduction}
Mass spectrometry is a key tool for life sciences and, in particular, for proteomics, the large-scale study of proteins \cite{Domon2006,Aebersold2016}. Proteins are essential for organisms and control mechanisms, such as signaling and metabolism \cite{Timp2020}. One of the most important properties of mass spectrometers is their sensitivity, which determines the minimal amount of a protein species needed to be detected. The sensitivity of conventional mass spectrometers is thousands of molecules \cite{Timp2020}. In comparison, NEMS-based mass spectrometers are able to detect a single molecule \cite{Hanay2012,Chaste2012}. However, nanomechanical resonators have an extremely low capture efficiency, so that it takes many molecules until one lands on the resonator for detection. To exploit the full potential of NEMS-based mass spectrometers, it is essential to reduce the sample loss. Besides an improved focusing of the molecular beam \cite{Dominguez-Medina2018}, this can be done by the arrangement of multiple resonators in a 2D array \cite{Sage2018,Bargatin2012}. The issue is that the standard electromechancial transduction methods do not allow the formation of dense arrays since electrical lines must be placed close to the resonators \cite{Tortonese1993,Truitt2007,Unterreithmeier2009,OConnell2010,Feng2008,Scheible2004,Schmid2016}. An exception is the transduction by surface acoustic waves (SAWs), which was recently demonstrated for the transduction of single mechanical pillar resonators \cite{Kahler2023}. The method is based on scattering of surface acoustic waves, which are launched and detected by interdigital transducers, distanced hundreds of micrometers away from the pillars.

Pillar resonators are well-suited for the sensing of masses and have two main advantages regarding mass spectrometry in comparison to horizontal mechanical structures: pillar resonators can be arranged in dense arrays due to their vertical structure and offer a defined landing position on the tip for the particles to be measured \cite{Paulitschke2019,Wasisto2013}. The latter enables a single mode operation of the pillars, instead of the 2-mode operation necessary for horizontal structures \cite{Schmid2010,Hanay2012}. 

Here, we demonstrate on the one hand the transduction of pillar pairs with very small separation distances of \SI{70}{\nano\meter} by SAWs. On the other hand, we show the SAW transduction of pillar pairs with a large separation distance of \SI{14.3}{\micro\meter}. Together, these results show the potential of SAWs to transduce dense arrays of pillar resonators, since in such arrays pillars can be closely and far separated. We measured the transmission scattering parameters of the overall devices as a function of frequency and modeled the results by a cascade of two port networks. In addition, we determine the response of the pillars to an added mass.   

\begin{figure}
  \begin{center}    
    \includegraphics{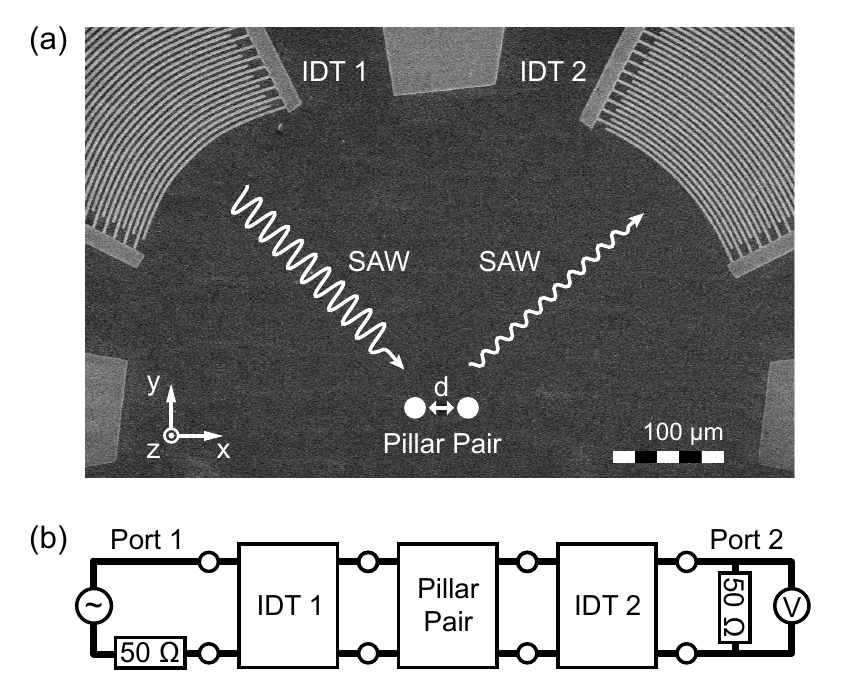}
    \caption{Transduction scheme. (a) Scanning electron microscope image of a device used in this study. The device consist of a piezoelectric substrate, two interdigital transducers (IDTs), and a pair of pillar resonators. For clarity, the pillars are represented by two white dots whose diameter is much larger then the actual pillar diameter of \SI{700}{\nano\meter}. The pillars are transduced by surface acoustic waves (SAWs) which are launched and detected by the IDTs, as illustrated by the white wavy lines. (b) Equivalent circuit model. The device is represented by a cascade of two-port networks, which are described by scattering parameters.}
    \label{Setup}
  \end{center}    
\end{figure}

\section{SAW transduction and pillar fabrication}
A scanning electron microscope image (SEM) of a device used in this study is shown in Fig.~\ref{Setup}(a). The substrate is $128^\circ$ Y-cut lithium niobate (LiNbO$_3$), which is piezoelectric. The piezoelectric effect is exploited to generate and detect SAWs by using interdigital transducers (IDTs). The IDTs are fabricated by photolithography and designed to focus Rayleigh-type SAWs. They have a central frequency of \SI{280}{\mega\hertz}, a bandwidth of around \SI{110}{\mega\hertz}, and each IDT covers an angle of $35^\circ$. In the focal point of the IDTs, we positioned either a single pillar resonator or a pair of pillars. The pillars are fabricated by Focused Electron Beam Induced Deposition (FEBID) using a platinum metal-organic precursor (MeCpPt$^\text{IV}$Me$_3$). The dimensions of the pillars were chosen so that their first compression mode falls within the frequency range of the IDTs \cite{Kahler2023}. All pillars had a diameter of \SI{700}{\nano\meter} and defined heights between \SI{2.3}{\micro\meter} to \SI{2.9}{\micro\meter}. However, the shape and material properties of different pillars can vary slighty, since FEBID is a very complex process. It involves a large number of independent variables, some of which are dynamically change during fabrication \cite{Winkler2018}. For this reason, we fabricated the two pillars of a pair simultaneously and not one pillar first and then the other.

The pillars are transduced by resonant scattering of SAWs, as illustrated in Fig.~\ref{Setup}(a). One of the IDTs emits a SAW and the other IDT measures the SAWs generated by the vibration of the pillars at resonance. We performed all measurements with a network analyzer and used on-wafer calibration. Further details to the device fabrication, the SAW transduction scheme and the experimental setup can be found in Ref.~\cite{Kahler2023}. 

\section{Equivalent circuit model}
It has been shown that the SAW transduction can be modelled by a cascade of two port networks described by scattering parameters, as illustrated in Fig.~\ref{Setup}(b) \cite{Kahler2023}. Since the IDTs are designed equivalently, the transmission scattering parameter of the whole device is given by 
\begin{equation}
S_\text{21}(f) =  C \, S_\text{IDT}^2(f) \, S_\text{P}(f) \; ,
\label{S21}
\end{equation}
where $f$ is the frequency of the applied input signal, $C$ is a constant, and $S_\text{IDT}$ and $S_\text{P}$ are the transmission scattering parameters of one of the IDTs and the pillar, respectively. In case of the transduction of a single pillar, the scattering parameter $S_\text{P}$ is given by \cite{Kahler2023}
\begin{equation}
S_\text{P}(f) = S_\text{eff} \, \frac{f^2}{f_0^2-f^2+\text{i} \, f \, \frac{f_0}{Q_0}} \; ,
\label{Sp1}
\end{equation}
where $S_\text{eff}$ is an effective scattering parameter, $f$ is the frequency of the incident SAW, and $f_0$ and $Q_0$ are the eigenfrequency and the total quality factor of the resonator, respectively.  

If multiple pillars are to be measured, they can not be placed all in the center. The consequences are additional phase differences between the measured signals of each pillar. First, the total distance traveled by the SAWs from one IDT to the other can differ for different pillar positions. Second, the average phase velocity of the SAWs traveled from one IDT to the other depends on the pillar position if the substrate material is anisotropic. For our devices, the phase velocities along the symmetry axes of the IDTs differ less than 10~\% \cite{Kovacs1990}. Hence, we assume that the additional phase differences between the pillars of our pillar pairs are small taking into account the orientation of the pillar pairs to the IDTs, as illustrated in Fig.~\ref{Setup}(a). 
If we assume that the additional phase differences between the signals of multiple pillars are small and the pillars do not couple, the transmision scattering parameter of the pillars $S_\text{P}$ is approximately given by 
\begin{equation}
S_\text{P}(f) \approx \sum_{n=1}^{N} S_{\text{eff},n} \, \frac{f^2}{f_n^2-f^2+\text{i} \, f \, \frac{f_n}{Q_n}} \; , 
\label{Sp2}
\end{equation}
where $N$ is the number of pillars or generally the number of uncoupled eigenmodes. Inserting Eq.~(\ref{Sp2}) into Eq.~(\ref{S21}) result in  
\begin{equation}
S_\text{21}(f) =  C \, S_\text{IDT}^2(f) \, \sum_{n=1}^{N} S_{\text{eff},n} \, \frac{f^2}{f_n^2-f^2+\text{i} \, f \, \frac{f_n}{Q_n}} \; .
\label{S21N}
\end{equation}

\section{Pillar pairs}
\begin{figure}
  \begin{center}    
    \includegraphics{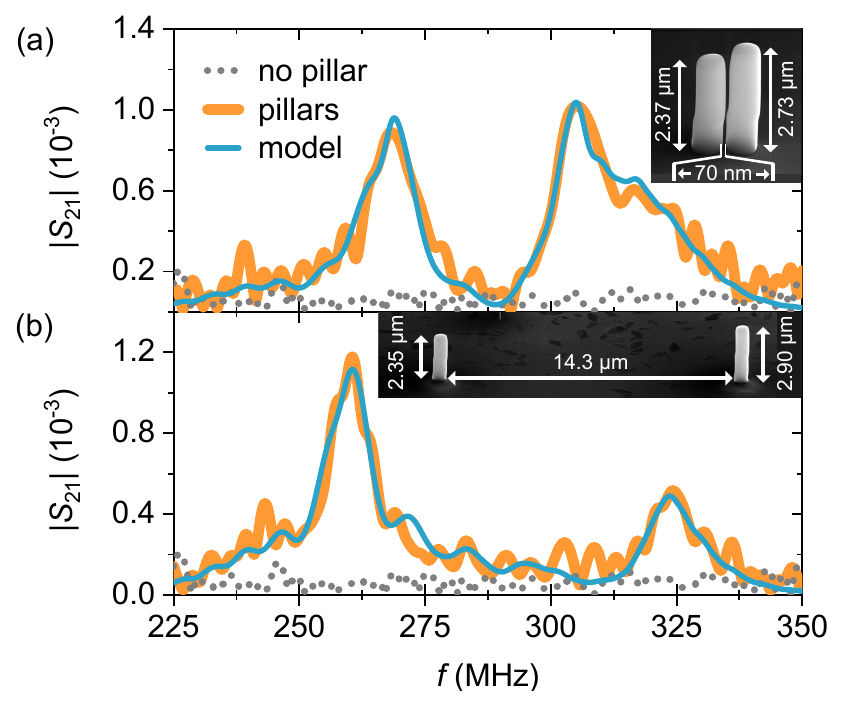}
    \caption{Frequency responses of two devices with closely and far separated pairs of mechanical pillar resonators. (a),(b) Amplitude of the measured transmission scattering parameter of the devices as a function of frequency. Scanning electron microscope images show the height of the measured pillars and their separation distance. The uncertainty in the pillars' height is around \SI{\pm50}{\nano\meter}. We fitted Eq.~(\ref{S21N}) for $N=2$ to the data. The model includes the measured frequency response of the two IDTs.}
    \label{Pairs}
  \end{center}    
\end{figure}

We measured the frequency response of two pairs of pillars and a device without any pillar as a reference. The pillar pairs were separated by \SI{70}{\nano\meter} and \SI{14.3}{\micro\meter}. The results of the measurements are shown in Fig.~\ref{Pairs} as well as SEM images of the pillar pairs. The pillars of each pair differed in height to distinguish them from one another. Two resonances are clearly visible for each pair. The resonances are well separated in frequency so that it can be assumed that the modes are not coupled \cite{Kahler2022}. To analyse the frequency response of the pairs, we first determined $S_\text{IDT}^2$ by measuring two focused IDTs facing each other (see Appendix) and then fitted Eq.~(\ref{S21N}) to the data for $N=2$ assuming that both pillars of a pair have the same quality factor $Q_0$. We fitted $|S_{21}|$ instead of $|S_{21}|/|S^2_\text{IDT}|$, since the normalization drastically increases the
noise outside the IDTs' frequency range. The fit of the model including the measured frequency response of the two IDTs is shown in Fig.~\ref{Pairs} and is in excellent agreement with the measured data of both pairs. 

In Table~\ref{TablePairs} the determined eigenfrequencies and quality factors of the pillars are listed. It can be seen that the ratios of the pillars' heights $h_2/h_1$ are equivalent to the inverse ratios of the their eigenfrequencies $f_1/f_2$. This indicates that we measured the first order compression mode of each pillar of a pair and not different eigenmodes of one of the pillars, since the eigenfrequency of a pillar's compression mode is inversely proportional to its height \cite{Schmid2006}. 
\setlength{\tabcolsep}{4pt}
\renewcommand{\arraystretch}{1.25}
\begin{table}[t]
\caption{Eigenfrequencies $f_n$ and quality factors $Q_0$ of two pillar pairs separated by a distance $d$. The pillars of a pair differed in height $h_n$. We determined $f_n$ and $Q_0$ by fitting Eq.~(\ref{S21N}) for $N=2$ to the frequency responses of the pillar pairs shown in Fig.~\ref{Pairs}. We assumed that both pillars of a pair have the same quality factor. The uncertainty in the pillars' height is around \SI{\pm50}{\nano\meter}.}
\centering
\begin{tabular}{ c c c  c c c c c }
 \hline
 \hline
 $d$ & $h_1$ & $f_1$& $h_2$ & $f_2$& \multirow{2}{1em}{$Q_0$} & \multirow{2}{2.5em}{$h_2/h_1$} & \multirow{2}{2.5em}{$f_1/f_2$} \\ 
 (\SI{}{\micro\meter}) & (\SI{}{\micro\meter}) & (\SI{}{\mega\hertz})  & (\SI{}{\micro\meter}) & (\SI{}{\mega\hertz}) & & &  \\
 \hline
 0.07 & 2.37 & 305 & 2.73 & 268 & 57 & 1.14 & 1.15  \\
14.3 & 2.35 & 322 & 2.90 & 259 & 33 & 1.23 & 1.24\\
 \hline
 \hline
\end{tabular}
\label{TablePairs}
\end{table}

\section{Frequency Response to Mass}
\begin{figure*}[t]    
  \begin{center}
    \includegraphics{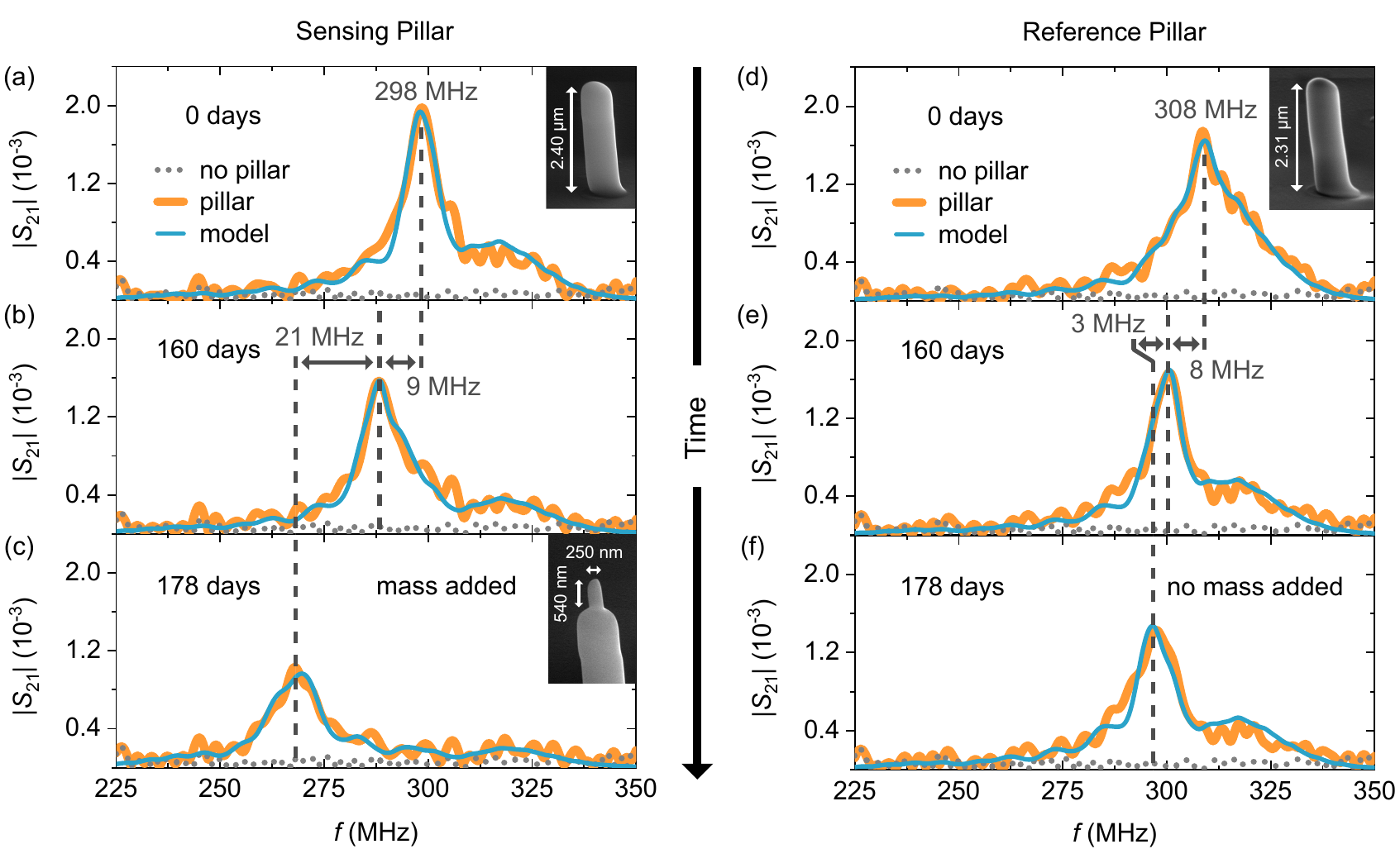}
    \caption{Response of a single pillar to added mass. Amplitude of the transmission scattering parameter of a device with a single pillar shortly after fabrication (a), after 160 days (b), and after the addition of an extra mass (c) as a function of frequency. (d)-(f) Amplitude of the transmission scattering parameter of a device with a reference pillar as a function of frequency. The pillar was placed close to the sensing pillar at all time, but no additional mass was added. Scanning electron microscope images show the height of the measured pillars and the dimensions of the added mass. The uncertainty in the height of the pillars and the added mass is around \SI{\pm50}{\nano\meter}. We fitted Eq.~(\ref{S21N}) for $N=1$ to the data. The model includes the measured frequency response of the two IDTs}
    \label{MassSensing}
  \end{center}    
\end{figure*}
We tested the platinum-carbon pillars for mass sensing applications and estimated the responsivity of the pillars to added mass. Previous studies have shown that the platinum-carbon material alters over time in ambient air \cite{Kolb2013,Arnold2018}. For this reason, we not only added a mass to a single pillar, but used a second pillar as reference, which was always kept close to the sensing pillar. Both pillars had similar dimensions and are depicted in Fig.~\ref{MassSensing}. It shows the frequency responses of the two pillars shortly after fabrication, after 160 days, and after the addition of an extra mass to one of the pillars (178 days). The sensing and the reference pillar had an eigenfrequency of \SI{298}{\mega\hertz} and \SI{308}{\mega\hertz} shortly after fabrication at 0~days, respectively, which we calculated by fitting Eq.~(\ref{S21N}) to the data for $N=1$. After 160~days, the eigenfrequency of both pillars decreased by a similar amount: \SI{-9}{\mega\hertz} and \SI{-8}{\mega\hertz}. Between 160~days and 178~days we added a mass to the sensing pillar, as shown in Fig.~\ref{MassSensing}(c), which resulted in a frequency decrease of \SI{-21}{\mega\hertz}. The reference pillar only showed a decrease by \SI{-3}{\mega\hertz} in the same time period. If we assume that the sensing pillar shows the same frequency drift over time as the refererence pillar, as it was nearly the case after 160~days, the added mass induced a frequency shift of roughly $\Delta f_0 = \SI{-18\pm2}{\mega\hertz}$. The added mass was deposited by FEBID using the same precursor as for the pillars. The deposited material is a mixture of platinum and carbon and has a density of around \SI{4.0\pm0.3}{\gram \,\cm^{-3}} \cite{Arnold2018}. The added mass had a diameter and height of \SI{250}{\nano\meter} and \SI{540\pm50}{\nano\meter}, respectively, which gives a total mass of $m_\text{add} = \SI{106\pm13}{\femto\gram}$. In comparison, the sensing pillar itself had a mass of $m_0 = \SI{3.7\pm0.3}{\pico\gram}$.

The fractional change of the eigenfrequency per added unit mass, which is called mass responsivity, is defined by \cite{Schmid2016}
\begin{equation}
R_\text{exp} = \frac{\Delta f_0}{m_\text{add}} \, \frac{1}{f_0} \; .
\label{ResMeas}
\end{equation}
We measured a responsivity for the sensing pillar of $R_\text{exp} = \SI{-588\pm98}{\nano\gram^{-1}}$ for $f_0 = \SI{289}{\mega\hertz}$. For a pillar vibrating in the first compression mode with a mass added at the top, the theoretical responsivity is given by \cite{Schmid2016}
\begin{equation}
R_\text{theo} \approx -\frac{2}{m_\text{0}} \; ,
\label{ResTheo}
\end{equation}
which result in $R_\text{theo} = \SI{-541\pm44}{\nano\gram^{-1}}$. It can be seen that the measured responsivity is in good agreement with the theoretical expected responsivity. From Eq.~(\ref{ResTheo}) it is clear that the responsivity can be improved by reducing the dimensions of the pillars.

\section{Conclusion}
We demonstrated the transduction of two pairs of pillars by surface acoustic waves. The pillar pairs were separated by \SI{70}{\nano\meter} and \SI{14.3}{\micro\meter}. We could actuate and detect the first compression mode of each pillar separately, which illustrates the potential of SAWs to transduce dense arrays of pillar resonators for mass spectrometry. After this successful proof of concept, the next step is to increase the number of pillar resonators, where coupling of the pillars becomes relevant due to the limited bandwidth of the IDTs. Coupling of the pillars complicates the interpretation of their frequency response, but also promises higher mass-sensitivity than individual resonators \cite{Spletzer2008}. Additionally, coupling of the pillars can also be used specifically to modify the quality factor of the pillars when it is dominated by radiation losses \cite{Kahler2022}. On our device, we can not place much more than four platinum-carbon pillars without a relevant coupling of the pillars. To increase the number of uncoupled resonators, we could increase the IDTs' bandwidth. However, a strong increase in the IDTs bandwidth result in a reduced transmission scattering parameter of the IDTs \cite{Morgan2007}. A more promising approach is to improve the resonators quality factor. Previous results suggest that the pillars' quality factor is dominated by intrinsic losses, since it seems to be independent of the pillars' dimensions or vibrational mode \cite{Kahler2023}. Hence, the quality factor of the pillars can be improved by using materials with lower intrinsic losses.

In addition to the transduction of the pillar pairs, we measured the mass responsivity of the pillars. However, our devices were not optimized for mass sensing applications. We operated at a central frequency of \SI{280}{\mega\hertz}, This is far away from the frequency limit of SAW devices, which is around several GHz. A modification of our devices to higher frequencies would allow us to transduce smaller pillars with higher responsivities, since the responsivity of the pillars is inversely proportional to their mass, as we discussed above.

\begin{acknowledgments}
We thank M. Buchholz for the fabrication of the IDT structures and R.G. West for many fruitful discussions. This work is supported by the European Research Council under the European Unions Horizon 2020 research and innovation program (Grant Agreement-716087-PLASMECS). The financial support by the Austrian Federal Ministry for Digital and Economic Affairs and the National Foundation for Research, Technology and Development is gratefully acknowledged (Christian Doppler Laboratory DEFINE).
\end{acknowledgments}

\section*{Contributions}
H.K. designed the samples, performed the electrical measurements, analysed the data, and wrote the original draft. R.W. fabricated the pillar resonators under the supervision of H.P. and supported the data analysis. H.A. supervised the electrical measurements and their analysis. H.K. wrote the paper with input from all authors. S.S. helped conceptualize and supervised the project. All authors reviewed and edited the manuscript.

\section*{Appendix: Transmission scattering parameter of the IDTs}
\begin{figure}[h]    
  \begin{center}
    \includegraphics{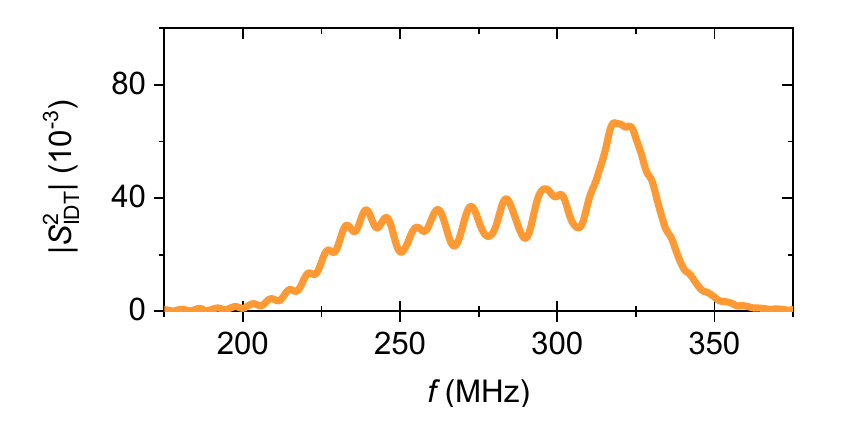}
    \caption{Frequency response of two interdigital transducers (IDTs) facing each other. The design of the IDTs is equivalent to the IDTs  shown in Fig.~\ref{Setup}(a).}
    \label{SIDT}
  \end{center}    
\end{figure}
We determined the transmission scattering parameter of the IDTs $S_\text{IDT}^2$ by measuring two focused IDTs facing each other. The IDTs were designed equivalently to the orthogonally arranged IDTs shown in Fig.~\ref{Setup}(a) and were placed along the crystallographic X-axis of the LiNbO$_3$ substrate. The result of the measurement is shown in Fig.~\ref{SIDT}.


\bibliography{PaperBibSAWTrans_V6}

\begin{thebibliography}{27}%
\makeatletter
\providecommand \@ifxundefined [1]{%
 \@ifx{#1\undefined}
}%
\providecommand \@ifnum [1]{%
 \ifnum #1\expandafter \@firstoftwo
 \else \expandafter \@secondoftwo
 \fi
}%
\providecommand \@ifx [1]{%
 \ifx #1\expandafter \@firstoftwo
 \else \expandafter \@secondoftwo
 \fi
}%
\providecommand \natexlab [1]{#1}%
\providecommand \enquote  [1]{``#1''}%
\providecommand \bibnamefont  [1]{#1}%
\providecommand \bibfnamefont [1]{#1}%
\providecommand \citenamefont [1]{#1}%
\providecommand \href@noop [0]{\@secondoftwo}%
\providecommand \href [0]{\begingroup \@sanitize@url \@href}%
\providecommand \@href[1]{\@@startlink{#1}\@@href}%
\providecommand \@@href[1]{\endgroup#1\@@endlink}%
\providecommand \@sanitize@url [0]{\catcode `\\12\catcode `\$12\catcode
  `\&12\catcode `\#12\catcode `\^12\catcode `\_12\catcode `\%12\relax}%
\providecommand \@@startlink[1]{}%
\providecommand \@@endlink[0]{}%
\providecommand \url  [0]{\begingroup\@sanitize@url \@url }%
\providecommand \@url [1]{\endgroup\@href {#1}{\urlprefix }}%
\providecommand \urlprefix  [0]{URL }%
\providecommand \Eprint [0]{\href }%
\providecommand \doibase [0]{http://dx.doi.org/}%
\providecommand \selectlanguage [0]{\@gobble}%
\providecommand \bibinfo  [0]{\@secondoftwo}%
\providecommand \bibfield  [0]{\@secondoftwo}%
\providecommand \translation [1]{[#1]}%
\providecommand \BibitemOpen [0]{}%
\providecommand \bibitemStop [0]{}%
\providecommand \bibitemNoStop [0]{.\EOS\space}%
\providecommand \EOS [0]{\spacefactor3000\relax}%
\providecommand \BibitemShut  [1]{\csname bibitem#1\endcsname}%
\let\auto@bib@innerbib\@empty
\bibitem [{\citenamefont {Domon}\ and\ \citenamefont
  {Aebersold}(2006)}]{Domon2006}%
  \BibitemOpen
  \bibfield  {author} {\bibinfo {author} {\bibfnamefont {B.}~\bibnamefont
  {Domon}}\ and\ \bibinfo {author} {\bibfnamefont {R.}~\bibnamefont
  {Aebersold}},\ }\href@noop {} {\bibfield  {journal} {\bibinfo  {journal}
  {Science}\ }\textbf {\bibinfo {volume} {312}},\ \bibinfo {pages} {212}
  (\bibinfo {year} {2006})}\BibitemShut {NoStop}%
\bibitem [{\citenamefont {Aebersold}\ and\ \citenamefont
  {Mann}(2016)}]{Aebersold2016}%
  \BibitemOpen
  \bibfield  {author} {\bibinfo {author} {\bibfnamefont {R.}~\bibnamefont
  {Aebersold}}\ and\ \bibinfo {author} {\bibfnamefont {M.}~\bibnamefont
  {Mann}},\ }\href@noop {} {\bibfield  {journal} {\bibinfo  {journal} {Nature}\
  }\textbf {\bibinfo {volume} {537}},\ \bibinfo {pages} {347} (\bibinfo {year}
  {2016})}\BibitemShut {NoStop}%
\bibitem [{\citenamefont {Timp}\ and\ \citenamefont {Timp}(2020)}]{Timp2020}%
  \BibitemOpen
  \bibfield  {author} {\bibinfo {author} {\bibfnamefont {W.}~\bibnamefont
  {Timp}}\ and\ \bibinfo {author} {\bibfnamefont {G.}~\bibnamefont {Timp}},\
  }\href@noop {} {\bibfield  {journal} {\bibinfo  {journal} {Science Advances}\
  }\textbf {\bibinfo {volume} {6}},\ \bibinfo {pages} {1} (\bibinfo {year}
  {2020})}\BibitemShut {NoStop}%
\bibitem [{\citenamefont {Hanay}\ \emph {et~al.}(2012)\citenamefont {Hanay},
  \citenamefont {Kelber}, \citenamefont {Naik}, \citenamefont {Chi},
  \citenamefont {Hentz}, \citenamefont {Bullard}, \citenamefont {Colinet},
  \citenamefont {Duraffourg},\ and\ \citenamefont {Roukes}}]{Hanay2012}%
  \BibitemOpen
  \bibfield  {author} {\bibinfo {author} {\bibfnamefont {M.~S.}\ \bibnamefont
  {Hanay}}, \bibinfo {author} {\bibfnamefont {S.}~\bibnamefont {Kelber}},
  \bibinfo {author} {\bibfnamefont {A.~K.}\ \bibnamefont {Naik}}, \bibinfo
  {author} {\bibfnamefont {D.}~\bibnamefont {Chi}}, \bibinfo {author}
  {\bibfnamefont {S.}~\bibnamefont {Hentz}}, \bibinfo {author} {\bibfnamefont
  {E.~C.}\ \bibnamefont {Bullard}}, \bibinfo {author} {\bibfnamefont
  {E.}~\bibnamefont {Colinet}}, \bibinfo {author} {\bibfnamefont
  {L.}~\bibnamefont {Duraffourg}}, \ and\ \bibinfo {author} {\bibfnamefont
  {M.~L.}\ \bibnamefont {Roukes}},\ }\href {\doibase 10.1038/nnano.2012.119}
  {\bibfield  {journal} {\bibinfo  {journal} {Nature Nanotechnology}\ }\textbf
  {\bibinfo {volume} {7}},\ \bibinfo {pages} {602} (\bibinfo {year}
  {2012})}\BibitemShut {NoStop}%
\bibitem [{\citenamefont {Chaste}\ \emph {et~al.}(2012)\citenamefont {Chaste},
  \citenamefont {Eichler}, \citenamefont {Moser}, \citenamefont {Ceballos},
  \citenamefont {Rurali},\ and\ \citenamefont {Bachtold}}]{Chaste2012}%
  \BibitemOpen
  \bibfield  {author} {\bibinfo {author} {\bibfnamefont {J.}~\bibnamefont
  {Chaste}}, \bibinfo {author} {\bibfnamefont {A.}~\bibnamefont {Eichler}},
  \bibinfo {author} {\bibfnamefont {J.}~\bibnamefont {Moser}}, \bibinfo
  {author} {\bibfnamefont {G.}~\bibnamefont {Ceballos}}, \bibinfo {author}
  {\bibfnamefont {R.}~\bibnamefont {Rurali}}, \ and\ \bibinfo {author}
  {\bibfnamefont {A.}~\bibnamefont {Bachtold}},\ }\href {\doibase
  10.1038/nnano.2012.42} {\bibfield  {journal} {\bibinfo  {journal} {Nature
  Nanotechnology}\ }\textbf {\bibinfo {volume} {7}},\ \bibinfo {pages} {301}
  (\bibinfo {year} {2012})}\BibitemShut {NoStop}%
\bibitem [{\citenamefont {Dominguez-Medina}\ \emph {et~al.}(2018)\citenamefont
  {Dominguez-Medina}, \citenamefont {Fostner}, \citenamefont {Defoort},
  \citenamefont {Sansa}, \citenamefont {Stark}, \citenamefont {Halim},
  \citenamefont {Vernhes}, \citenamefont {Gely}, \citenamefont {Jourdan},
  \citenamefont {Alava}, \citenamefont {Boulanger}, \citenamefont {Masselon},\
  and\ \citenamefont {Hentz}}]{Dominguez-Medina2018}%
  \BibitemOpen
  \bibfield  {author} {\bibinfo {author} {\bibfnamefont {S.}~\bibnamefont
  {Dominguez-Medina}}, \bibinfo {author} {\bibfnamefont {S.}~\bibnamefont
  {Fostner}}, \bibinfo {author} {\bibfnamefont {M.}~\bibnamefont {Defoort}},
  \bibinfo {author} {\bibfnamefont {M.}~\bibnamefont {Sansa}}, \bibinfo
  {author} {\bibfnamefont {A.~K.}\ \bibnamefont {Stark}}, \bibinfo {author}
  {\bibfnamefont {M.~A.}\ \bibnamefont {Halim}}, \bibinfo {author}
  {\bibfnamefont {E.}~\bibnamefont {Vernhes}}, \bibinfo {author} {\bibfnamefont
  {M.}~\bibnamefont {Gely}}, \bibinfo {author} {\bibfnamefont {G.}~\bibnamefont
  {Jourdan}}, \bibinfo {author} {\bibfnamefont {T.}~\bibnamefont {Alava}},
  \bibinfo {author} {\bibfnamefont {P.}~\bibnamefont {Boulanger}}, \bibinfo
  {author} {\bibfnamefont {C.}~\bibnamefont {Masselon}}, \ and\ \bibinfo
  {author} {\bibfnamefont {S.}~\bibnamefont {Hentz}},\ }\href {\doibase
  10.1126/science.aat6457} {\bibfield  {journal} {\bibinfo  {journal}
  {Science}\ }\textbf {\bibinfo {volume} {362}},\ \bibinfo {pages} {918}
  (\bibinfo {year} {2018})}\BibitemShut {NoStop}%
\bibitem [{\citenamefont {Sage}\ \emph {et~al.}(2018)\citenamefont {Sage},
  \citenamefont {Sansa}, \citenamefont {Fostner}, \citenamefont {Defoort},
  \citenamefont {G{\'{e}}ly}, \citenamefont {Naik}, \citenamefont {Morel},
  \citenamefont {Duraffourg}, \citenamefont {Roukes}, \citenamefont {Alava},
  \citenamefont {Jourdan}, \citenamefont {Colinet}, \citenamefont {Masselon},
  \citenamefont {Brenac},\ and\ \citenamefont {Hentz}}]{Sage2018}%
  \BibitemOpen
  \bibfield  {author} {\bibinfo {author} {\bibfnamefont {E.}~\bibnamefont
  {Sage}}, \bibinfo {author} {\bibfnamefont {M.}~\bibnamefont {Sansa}},
  \bibinfo {author} {\bibfnamefont {S.}~\bibnamefont {Fostner}}, \bibinfo
  {author} {\bibfnamefont {M.}~\bibnamefont {Defoort}}, \bibinfo {author}
  {\bibfnamefont {M.}~\bibnamefont {G{\'{e}}ly}}, \bibinfo {author}
  {\bibfnamefont {A.~K.}\ \bibnamefont {Naik}}, \bibinfo {author}
  {\bibfnamefont {R.}~\bibnamefont {Morel}}, \bibinfo {author} {\bibfnamefont
  {L.}~\bibnamefont {Duraffourg}}, \bibinfo {author} {\bibfnamefont {M.~L.}\
  \bibnamefont {Roukes}}, \bibinfo {author} {\bibfnamefont {T.}~\bibnamefont
  {Alava}}, \bibinfo {author} {\bibfnamefont {G.}~\bibnamefont {Jourdan}},
  \bibinfo {author} {\bibfnamefont {E.}~\bibnamefont {Colinet}}, \bibinfo
  {author} {\bibfnamefont {C.}~\bibnamefont {Masselon}}, \bibinfo {author}
  {\bibfnamefont {A.}~\bibnamefont {Brenac}}, \ and\ \bibinfo {author}
  {\bibfnamefont {S.}~\bibnamefont {Hentz}},\ }\href {\doibase
  10.1038/s41467-018-05783-4} {\bibfield  {journal} {\bibinfo  {journal}
  {Nature Communications}\ }\textbf {\bibinfo {volume} {9}},\ \bibinfo {pages}
  {1} (\bibinfo {year} {2018})}\BibitemShut {NoStop}%
\bibitem [{\citenamefont {Bargatin}\ \emph {et~al.}(2012)\citenamefont
  {Bargatin}, \citenamefont {Myers}, \citenamefont {Aldridge}, \citenamefont
  {Marcoux}, \citenamefont {Brianceau}, \citenamefont {Duraffourg},
  \citenamefont {Colinet}, \citenamefont {Hentz}, \citenamefont {Andreucci},\
  and\ \citenamefont {Roukes}}]{Bargatin2012}%
  \BibitemOpen
  \bibfield  {author} {\bibinfo {author} {\bibfnamefont {I.}~\bibnamefont
  {Bargatin}}, \bibinfo {author} {\bibfnamefont {E.~B.}\ \bibnamefont {Myers}},
  \bibinfo {author} {\bibfnamefont {J.~S.}\ \bibnamefont {Aldridge}}, \bibinfo
  {author} {\bibfnamefont {C.}~\bibnamefont {Marcoux}}, \bibinfo {author}
  {\bibfnamefont {P.}~\bibnamefont {Brianceau}}, \bibinfo {author}
  {\bibfnamefont {L.}~\bibnamefont {Duraffourg}}, \bibinfo {author}
  {\bibfnamefont {E.}~\bibnamefont {Colinet}}, \bibinfo {author} {\bibfnamefont
  {S.}~\bibnamefont {Hentz}}, \bibinfo {author} {\bibfnamefont
  {P.}~\bibnamefont {Andreucci}}, \ and\ \bibinfo {author} {\bibfnamefont
  {M.~L.}\ \bibnamefont {Roukes}},\ }\href {\doibase 10.1021/nl2037479}
  {\bibfield  {journal} {\bibinfo  {journal} {Nano Letters}\ }\textbf {\bibinfo
  {volume} {12}},\ \bibinfo {pages} {1269} (\bibinfo {year}
  {2012})}\BibitemShut {NoStop}%
\bibitem [{\citenamefont {Tortonese}\ \emph {et~al.}(1993)\citenamefont
  {Tortonese}, \citenamefont {Barrett},\ and\ \citenamefont
  {Quate}}]{Tortonese1993}%
  \BibitemOpen
  \bibfield  {author} {\bibinfo {author} {\bibfnamefont {M.}~\bibnamefont
  {Tortonese}}, \bibinfo {author} {\bibfnamefont {R.~C.}\ \bibnamefont
  {Barrett}}, \ and\ \bibinfo {author} {\bibfnamefont {C.~F.}\ \bibnamefont
  {Quate}},\ }\href {\doibase 10.1063/1.108593} {\bibfield  {journal} {\bibinfo
   {journal} {Applied Physics Letters}\ }\textbf {\bibinfo {volume} {62}},\
  \bibinfo {pages} {834} (\bibinfo {year} {1993})}\BibitemShut {NoStop}%
\bibitem [{\citenamefont {Truitt}\ \emph {et~al.}(2007)\citenamefont {Truitt},
  \citenamefont {Hertzberg}, \citenamefont {Huang}, \citenamefont {Ekinci},\
  and\ \citenamefont {Schwab}}]{Truitt2007}%
  \BibitemOpen
  \bibfield  {author} {\bibinfo {author} {\bibfnamefont {P.~A.}\ \bibnamefont
  {Truitt}}, \bibinfo {author} {\bibfnamefont {J.~B.}\ \bibnamefont
  {Hertzberg}}, \bibinfo {author} {\bibfnamefont {C.~C.}\ \bibnamefont
  {Huang}}, \bibinfo {author} {\bibfnamefont {K.~L.}\ \bibnamefont {Ekinci}}, \
  and\ \bibinfo {author} {\bibfnamefont {K.~C.}\ \bibnamefont {Schwab}},\
  }\href {\doibase 10.1021/nl062278g} {\bibfield  {journal} {\bibinfo
  {journal} {Nano Letters}\ }\textbf {\bibinfo {volume} {7}},\ \bibinfo {pages}
  {120} (\bibinfo {year} {2007})}\BibitemShut {NoStop}%
\bibitem [{\citenamefont {Unterreithmeier}\ \emph {et~al.}(2009)\citenamefont
  {Unterreithmeier}, \citenamefont {Weig},\ and\ \citenamefont
  {Kotthaus}}]{Unterreithmeier2009}%
  \BibitemOpen
  \bibfield  {author} {\bibinfo {author} {\bibfnamefont {Q.~P.}\ \bibnamefont
  {Unterreithmeier}}, \bibinfo {author} {\bibfnamefont {E.~M.}\ \bibnamefont
  {Weig}}, \ and\ \bibinfo {author} {\bibfnamefont {J.~P.}\ \bibnamefont
  {Kotthaus}},\ }\href {\doibase 10.1038/nature07932} {\bibfield  {journal}
  {\bibinfo  {journal} {Nature}\ }\textbf {\bibinfo {volume} {458}},\ \bibinfo
  {pages} {1001} (\bibinfo {year} {2009})}\BibitemShut {NoStop}%
\bibitem [{\citenamefont {O'Connell}\ \emph {et~al.}(2010)\citenamefont
  {O'Connell}, \citenamefont {Hofheinz}, \citenamefont {Ansmann}, \citenamefont
  {Bialczak}, \citenamefont {Lenander}, \citenamefont {Lucero}, \citenamefont
  {Neeley}, \citenamefont {Sank}, \citenamefont {Wang}, \citenamefont {Weides},
  \citenamefont {Wenner}, \citenamefont {Martinis},\ and\ \citenamefont
  {Cleland}}]{OConnell2010}%
  \BibitemOpen
  \bibfield  {author} {\bibinfo {author} {\bibfnamefont {A.~D.}\ \bibnamefont
  {O'Connell}}, \bibinfo {author} {\bibfnamefont {M.}~\bibnamefont {Hofheinz}},
  \bibinfo {author} {\bibfnamefont {M.}~\bibnamefont {Ansmann}}, \bibinfo
  {author} {\bibfnamefont {R.~C.}\ \bibnamefont {Bialczak}}, \bibinfo {author}
  {\bibfnamefont {M.}~\bibnamefont {Lenander}}, \bibinfo {author}
  {\bibfnamefont {E.}~\bibnamefont {Lucero}}, \bibinfo {author} {\bibfnamefont
  {M.}~\bibnamefont {Neeley}}, \bibinfo {author} {\bibfnamefont
  {D.}~\bibnamefont {Sank}}, \bibinfo {author} {\bibfnamefont {H.}~\bibnamefont
  {Wang}}, \bibinfo {author} {\bibfnamefont {M.}~\bibnamefont {Weides}},
  \bibinfo {author} {\bibfnamefont {J.}~\bibnamefont {Wenner}}, \bibinfo
  {author} {\bibfnamefont {J.~M.}\ \bibnamefont {Martinis}}, \ and\ \bibinfo
  {author} {\bibfnamefont {A.~N.}\ \bibnamefont {Cleland}},\ }\href {\doibase
  10.1038/nature08967} {\bibfield  {journal} {\bibinfo  {journal} {Nature}\
  }\textbf {\bibinfo {volume} {464}},\ \bibinfo {pages} {697} (\bibinfo {year}
  {2010})}\BibitemShut {NoStop}%
\bibitem [{\citenamefont {Feng}\ \emph {et~al.}(2008)\citenamefont {Feng},
  \citenamefont {White}, \citenamefont {Hajimiri},\ and\ \citenamefont
  {Roukes}}]{Feng2008}%
  \BibitemOpen
  \bibfield  {author} {\bibinfo {author} {\bibfnamefont {X.~L.}\ \bibnamefont
  {Feng}}, \bibinfo {author} {\bibfnamefont {C.~J.}\ \bibnamefont {White}},
  \bibinfo {author} {\bibfnamefont {A.}~\bibnamefont {Hajimiri}}, \ and\
  \bibinfo {author} {\bibfnamefont {M.~L.}\ \bibnamefont {Roukes}},\ }\href
  {\doibase 10.1038/nnano.2008.125} {\bibfield  {journal} {\bibinfo  {journal}
  {Nature Nanotechnology}\ }\textbf {\bibinfo {volume} {3}},\ \bibinfo {pages}
  {342} (\bibinfo {year} {2008})}\BibitemShut {NoStop}%
\bibitem [{\citenamefont {Scheible}\ and\ \citenamefont
  {Blick}(2004)}]{Scheible2004}%
  \BibitemOpen
  \bibfield  {author} {\bibinfo {author} {\bibfnamefont {D.~V.}\ \bibnamefont
  {Scheible}}\ and\ \bibinfo {author} {\bibfnamefont {R.~H.}\ \bibnamefont
  {Blick}},\ }\href {\doibase 10.1063/1.1759371} {\bibfield  {journal}
  {\bibinfo  {journal} {Applied Physics Letters}\ }\textbf {\bibinfo {volume}
  {84}},\ \bibinfo {pages} {4632} (\bibinfo {year} {2004})}\BibitemShut
  {NoStop}%
\bibitem [{\citenamefont {Schmid}\ \emph {et~al.}(2016)\citenamefont {Schmid},
  \citenamefont {Villanueva},\ and\ \citenamefont {Roukes}}]{Schmid2016}%
  \BibitemOpen
  \bibfield  {author} {\bibinfo {author} {\bibfnamefont {S.}~\bibnamefont
  {Schmid}}, \bibinfo {author} {\bibfnamefont {L.~G.}\ \bibnamefont
  {Villanueva}}, \ and\ \bibinfo {author} {\bibfnamefont {M.~L.}\ \bibnamefont
  {Roukes}},\ }\href@noop {} {\emph {\bibinfo {title} {{Fundamentals of
  nanomechanical resonators}}}}\ (\bibinfo  {publisher} {Springer},\ \bibinfo
  {year} {2016})\BibitemShut {NoStop}%
\bibitem [{\citenamefont {K{\"{a}}hler}\ \emph {et~al.}(2023)\citenamefont
  {K{\"{a}}hler}, \citenamefont {Arthaber}, \citenamefont {Winkler},
  \citenamefont {West}, \citenamefont {Ignat}, \citenamefont {Plank},\ and\
  \citenamefont {Schmid}}]{Kahler2023}%
  \BibitemOpen
  \bibfield  {author} {\bibinfo {author} {\bibfnamefont {H.}~\bibnamefont
  {K{\"{a}}hler}}, \bibinfo {author} {\bibfnamefont {H.}~\bibnamefont
  {Arthaber}}, \bibinfo {author} {\bibfnamefont {R.}~\bibnamefont {Winkler}},
  \bibinfo {author} {\bibfnamefont {R.~G.}\ \bibnamefont {West}}, \bibinfo
  {author} {\bibfnamefont {I.}~\bibnamefont {Ignat}}, \bibinfo {author}
  {\bibfnamefont {H.}~\bibnamefont {Plank}}, \ and\ \bibinfo {author}
  {\bibfnamefont {S.}~\bibnamefont {Schmid}},\ }\href@noop {} {\bibfield
  {journal} {\bibinfo  {journal} {arXiv preprint arXiv:2210.09069}\ } (\bibinfo
  {year} {2023})}\BibitemShut {NoStop}%
\bibitem [{\citenamefont {Paulitschke}\ \emph {et~al.}(2019)\citenamefont
  {Paulitschke}, \citenamefont {Keber}, \citenamefont {Lebedev}, \citenamefont
  {Stephan}, \citenamefont {Lorenz}, \citenamefont {Hasselmann}, \citenamefont
  {Heinrich},\ and\ \citenamefont {Weig}}]{Paulitschke2019}%
  \BibitemOpen
  \bibfield  {author} {\bibinfo {author} {\bibfnamefont {P.}~\bibnamefont
  {Paulitschke}}, \bibinfo {author} {\bibfnamefont {F.}~\bibnamefont {Keber}},
  \bibinfo {author} {\bibfnamefont {A.}~\bibnamefont {Lebedev}}, \bibinfo
  {author} {\bibfnamefont {J.}~\bibnamefont {Stephan}}, \bibinfo {author}
  {\bibfnamefont {H.}~\bibnamefont {Lorenz}}, \bibinfo {author} {\bibfnamefont
  {S.}~\bibnamefont {Hasselmann}}, \bibinfo {author} {\bibfnamefont
  {D.}~\bibnamefont {Heinrich}}, \ and\ \bibinfo {author} {\bibfnamefont
  {E.~M.}\ \bibnamefont {Weig}},\ }\href {\doibase
  10.1021/acs.nanolett.8b02568} {\bibfield  {journal} {\bibinfo  {journal}
  {Nano Letters}\ }\textbf {\bibinfo {volume} {19}},\ \bibinfo {pages} {2207}
  (\bibinfo {year} {2019})}\BibitemShut {NoStop}%
\bibitem [{\citenamefont {Wasisto}\ \emph {et~al.}(2013)\citenamefont
  {Wasisto}, \citenamefont {Merzsch}, \citenamefont {Stranz}, \citenamefont
  {Waag}, \citenamefont {Uhde}, \citenamefont {Salthammer},\ and\ \citenamefont
  {Peiner}}]{Wasisto2013}%
  \BibitemOpen
  \bibfield  {author} {\bibinfo {author} {\bibfnamefont {H.~S.}\ \bibnamefont
  {Wasisto}}, \bibinfo {author} {\bibfnamefont {S.}~\bibnamefont {Merzsch}},
  \bibinfo {author} {\bibfnamefont {A.}~\bibnamefont {Stranz}}, \bibinfo
  {author} {\bibfnamefont {A.}~\bibnamefont {Waag}}, \bibinfo {author}
  {\bibfnamefont {E.}~\bibnamefont {Uhde}}, \bibinfo {author} {\bibfnamefont
  {T.}~\bibnamefont {Salthammer}}, \ and\ \bibinfo {author} {\bibfnamefont
  {E.}~\bibnamefont {Peiner}},\ }\href {\doibase 10.1016/j.snb.2013.02.053}
  {\bibfield  {journal} {\bibinfo  {journal} {Sensors and Actuators, B:
  Chemical}\ }\textbf {\bibinfo {volume} {189}},\ \bibinfo {pages} {146}
  (\bibinfo {year} {2013})}\BibitemShut {NoStop}%
\bibitem [{\citenamefont {Schmid}\ \emph {et~al.}(2010)\citenamefont {Schmid},
  \citenamefont {Dohn},\ and\ \citenamefont {Boisen}}]{Schmid2010}%
  \BibitemOpen
  \bibfield  {author} {\bibinfo {author} {\bibfnamefont {S.}~\bibnamefont
  {Schmid}}, \bibinfo {author} {\bibfnamefont {S.}~\bibnamefont {Dohn}}, \ and\
  \bibinfo {author} {\bibfnamefont {A.}~\bibnamefont {Boisen}},\ }\href
  {\doibase 10.3390/s100908092} {\bibfield  {journal} {\bibinfo  {journal}
  {Sensors}\ }\textbf {\bibinfo {volume} {10}},\ \bibinfo {pages} {8192}
  (\bibinfo {year} {2010})}\BibitemShut {NoStop}%
\bibitem [{\citenamefont {Winkler}\ \emph {et~al.}(2018)\citenamefont
  {Winkler}, \citenamefont {Lewis}, \citenamefont {Fowlkes}, \citenamefont
  {Rack},\ and\ \citenamefont {Plank}}]{Winkler2018}%
  \BibitemOpen
  \bibfield  {author} {\bibinfo {author} {\bibfnamefont {R.}~\bibnamefont
  {Winkler}}, \bibinfo {author} {\bibfnamefont {B.~B.}\ \bibnamefont {Lewis}},
  \bibinfo {author} {\bibfnamefont {J.~D.}\ \bibnamefont {Fowlkes}}, \bibinfo
  {author} {\bibfnamefont {P.~D.}\ \bibnamefont {Rack}}, \ and\ \bibinfo
  {author} {\bibfnamefont {H.}~\bibnamefont {Plank}},\ }\href {\doibase
  10.1021/acsanm.8b00158} {\bibfield  {journal} {\bibinfo  {journal} {ACS
  Applied Nano Materials}\ }\textbf {\bibinfo {volume} {1}},\ \bibinfo {pages}
  {1014} (\bibinfo {year} {2018})}\BibitemShut {NoStop}%
\bibitem [{\citenamefont {Kovacs}\ \emph {et~al.}(1990)\citenamefont {Kovacs},
  \citenamefont {Anhorn}, \citenamefont {Engan}, \citenamefont {Visintini},\
  and\ \citenamefont {Ruppel}}]{Kovacs1990}%
  \BibitemOpen
  \bibfield  {author} {\bibinfo {author} {\bibfnamefont {G.}~\bibnamefont
  {Kovacs}}, \bibinfo {author} {\bibfnamefont {M.}~\bibnamefont {Anhorn}},
  \bibinfo {author} {\bibfnamefont {H.}~\bibnamefont {Engan}}, \bibinfo
  {author} {\bibfnamefont {G.}~\bibnamefont {Visintini}}, \ and\ \bibinfo
  {author} {\bibfnamefont {C.}~\bibnamefont {Ruppel}},\ }\href@noop {}
  {\bibfield  {journal} {\bibinfo  {journal} {IEEE Symposium on Ultrasonics}\
  ,\ \bibinfo {pages} {435}} (\bibinfo {year} {1990})}\BibitemShut {NoStop}%
\bibitem [{\citenamefont {K{\"{a}}hler}\ \emph {et~al.}(2022)\citenamefont
  {K{\"{a}}hler}, \citenamefont {Platz},\ and\ \citenamefont
  {Schmid}}]{Kahler2022}%
  \BibitemOpen
  \bibfield  {author} {\bibinfo {author} {\bibfnamefont {H.}~\bibnamefont
  {K{\"{a}}hler}}, \bibinfo {author} {\bibfnamefont {D.}~\bibnamefont {Platz}},
  \ and\ \bibinfo {author} {\bibfnamefont {S.}~\bibnamefont {Schmid}},\ }\href
  {\doibase 10.1038/s42005-022-00895-2} {\bibfield  {journal} {\bibinfo
  {journal} {Communications Physics}\ }\textbf {\bibinfo {volume} {5}},\
  \bibinfo {pages} {1} (\bibinfo {year} {2022})}\BibitemShut {NoStop}%
\bibitem [{\citenamefont {Schmid}\ \emph {et~al.}(2006)\citenamefont {Schmid},
  \citenamefont {Wendlandt}, \citenamefont {Junker},\ and\ \citenamefont
  {Hierold}}]{Schmid2006}%
  \BibitemOpen
  \bibfield  {author} {\bibinfo {author} {\bibfnamefont {S.}~\bibnamefont
  {Schmid}}, \bibinfo {author} {\bibfnamefont {M.}~\bibnamefont {Wendlandt}},
  \bibinfo {author} {\bibfnamefont {D.}~\bibnamefont {Junker}}, \ and\ \bibinfo
  {author} {\bibfnamefont {C.}~\bibnamefont {Hierold}},\ }\href {\doibase
  10.1063/1.2362590} {\bibfield  {journal} {\bibinfo  {journal} {Applied
  Physics Letters}\ }\textbf {\bibinfo {volume} {89}},\ \bibinfo {pages} {1}
  (\bibinfo {year} {2006})}\BibitemShut {NoStop}%
\bibitem [{\citenamefont {Kolb}\ \emph {et~al.}(2013)\citenamefont {Kolb},
  \citenamefont {Schmoltner}, \citenamefont {Huth}, \citenamefont {Hohenau},
  \citenamefont {Krenn}, \citenamefont {Klug}, \citenamefont {List},\ and\
  \citenamefont {Plank}}]{Kolb2013}%
  \BibitemOpen
  \bibfield  {author} {\bibinfo {author} {\bibfnamefont {F.}~\bibnamefont
  {Kolb}}, \bibinfo {author} {\bibfnamefont {K.}~\bibnamefont {Schmoltner}},
  \bibinfo {author} {\bibfnamefont {M.}~\bibnamefont {Huth}}, \bibinfo {author}
  {\bibfnamefont {A.}~\bibnamefont {Hohenau}}, \bibinfo {author} {\bibfnamefont
  {J.}~\bibnamefont {Krenn}}, \bibinfo {author} {\bibfnamefont
  {A.}~\bibnamefont {Klug}}, \bibinfo {author} {\bibfnamefont {E.~J.}\
  \bibnamefont {List}}, \ and\ \bibinfo {author} {\bibfnamefont
  {H.}~\bibnamefont {Plank}},\ }\href {\doibase 10.1088/0957-4484/24/30/305501}
  {\bibfield  {journal} {\bibinfo  {journal} {Nanotechnology}\ }\textbf
  {\bibinfo {volume} {24}},\ \bibinfo {pages} {1} (\bibinfo {year}
  {2013})}\BibitemShut {NoStop}%
\bibitem [{\citenamefont {Arnold}\ \emph {et~al.}(2018)\citenamefont {Arnold},
  \citenamefont {Winkler}, \citenamefont {Stermitz}, \citenamefont {Orthacker},
  \citenamefont {Noh}, \citenamefont {Fowlkes}, \citenamefont {Kothleitner},
  \citenamefont {Huth}, \citenamefont {Rack},\ and\ \citenamefont
  {Plank}}]{Arnold2018}%
  \BibitemOpen
  \bibfield  {author} {\bibinfo {author} {\bibfnamefont {G.}~\bibnamefont
  {Arnold}}, \bibinfo {author} {\bibfnamefont {R.}~\bibnamefont {Winkler}},
  \bibinfo {author} {\bibfnamefont {M.}~\bibnamefont {Stermitz}}, \bibinfo
  {author} {\bibfnamefont {A.}~\bibnamefont {Orthacker}}, \bibinfo {author}
  {\bibfnamefont {J.~H.}\ \bibnamefont {Noh}}, \bibinfo {author} {\bibfnamefont
  {J.~D.}\ \bibnamefont {Fowlkes}}, \bibinfo {author} {\bibfnamefont
  {G.}~\bibnamefont {Kothleitner}}, \bibinfo {author} {\bibfnamefont
  {M.}~\bibnamefont {Huth}}, \bibinfo {author} {\bibfnamefont {P.~D.}\
  \bibnamefont {Rack}}, \ and\ \bibinfo {author} {\bibfnamefont
  {H.}~\bibnamefont {Plank}},\ }\href {\doibase 10.1002/adfm.201707387}
  {\bibfield  {journal} {\bibinfo  {journal} {Advanced Functional Materials}\
  }\textbf {\bibinfo {volume} {28}},\ \bibinfo {pages} {1} (\bibinfo {year}
  {2018})}\BibitemShut {NoStop}%
\bibitem [{\citenamefont {Spletzer}\ \emph {et~al.}(2008)\citenamefont
  {Spletzer}, \citenamefont {Raman}, \citenamefont {Sumali},\ and\
  \citenamefont {Sullivan}}]{Spletzer2008}%
  \BibitemOpen
  \bibfield  {author} {\bibinfo {author} {\bibfnamefont {M.}~\bibnamefont
  {Spletzer}}, \bibinfo {author} {\bibfnamefont {A.}~\bibnamefont {Raman}},
  \bibinfo {author} {\bibfnamefont {H.}~\bibnamefont {Sumali}}, \ and\ \bibinfo
  {author} {\bibfnamefont {J.~P.}\ \bibnamefont {Sullivan}},\ }\href@noop {}
  {\bibfield  {journal} {\bibinfo  {journal} {Applied Physics Letters}\
  }\textbf {\bibinfo {volume} {92}},\ \bibinfo {pages} {1} (\bibinfo {year}
  {2008})}\BibitemShut {NoStop}%
\bibitem [{\citenamefont {Morgan}(2007)}]{Morgan2007}%
  \BibitemOpen
  \bibfield  {author} {\bibinfo {author} {\bibfnamefont {D.}~\bibnamefont
  {Morgan}},\ }\href@noop {} {\emph {\bibinfo {title} {Surface Acoustic Wave
  Filters}}},\ \bibinfo {edition} {2nd}\ ed.\ (\bibinfo  {publisher} {Elsevier
  Ltd.},\ \bibinfo {year} {2007})\BibitemShut {NoStop}%
\end{thebibliography}%

\end{document}